# Unbreakable cryptosystem for common people

**Arindam Mitra**

**Anushakti Abasan, Uttar Falguni -7, 1/AF, Salt lake,
Kolkata, West Bengal, 700064, India.**

Abstract: It has been found that an algorithm can generate true random numbers on classical computer. The algorithm can be used to generate unbreakable message PIN (personal identification number) and password.

Cryptography has long history [1]. Interestingly, it has been found that taking key material from open source it is possible to design unbreakable crypto-system. Following the technique [2] an algorithm [3] has also been designed for the generation of arbitrary numbers of random numbers. Using the algorithm [3] distant parties can generate common key if they give the same inputs which are three sequences of random numbers. It means distant parties have to secretly share the initial inputs. Common people can use the algorithms because only for one time common people can easily generate some random numbers by tossing coin. No need of taking any help from third party to procure secret key.

Suppose the output of the algorithm, that is a n-bit key, is used to encrypt a n-bit message. Eavesdropper's probability of guessing the n-bit key/message is $1/2^n$. The system is information theoretically completely secure. However, eavesdropper can steal some messages and decode the message encrypting keys to know their secret keys $S_0^M$, $S_1^F(S_2^F)$ and $S^C$, and thereby unused message encrypting keys. This attack is known as plaintext-attack. To know this information eavesdropper has to generate outputs giving his own input sequences $S_0^M$, $S_1^F(S_2^F)$ and $S^C$. Total number of probable three input sequences is $2^{2n}(2^{2n}-1)(2^{2n}-2)$. Practically this is impossible to check even for small n, say $= 500$ if we keep in mind that the estimated number of atoms in this universe is approximately $2^{500}$. Plain-text attack can still be foiled following the same technique described for the previous scheme [2]. After encrypting some independent keys in the generated keys one party has to transmit the independent keys as secret message to other party. The independent shared keys can be used both in message encryption and authentication purpose. That is, the algorithm can generate unbreakable message, PIN and password.

It is easy to see that the algorithm can also be used as N-digit crypto-system. Instead of initially secretly sharing three sequences of random numbers parties can secretly share a N-digit number to generate random numbers. There are many ways N-digit crypto-system can be designed. We shall present the best design in the next version.